\documentclass[prl,floatfix,showpacs,twocolumn,tightenlines,nofootinbib,superscriptaddress]{revtex4-1}

\usepackage[utf8]{inputenc}
\usepackage[T1]{fontenc}
\usepackage{braket}
\usepackage{graphicx}
\usepackage{booktabs}
\usepackage{mathtools}	
\usepackage{amssymb}
\usepackage{amsmath}		
\usepackage{amsfonts}
\usepackage{upgreek}
\usepackage{mathrsfs}
\usepackage{color}
\usepackage{amsthm}
\usepackage{psfrag}
\usepackage{ifsym}
\usepackage{dsfont}
\usepackage{enumerate}
\usepackage{enumitem}
\usepackage{multirow} 
\usepackage{hyperref}
\hypersetup{colorlinks=true,linkcolor=red,citecolor=blue}
\usepackage{float}
\usepackage[separate-uncertainty=true]{siunitx}
\usepackage{setspace}
\usepackage[sort&compress]{natbib}
\usepackage[dvipsnames]{xcolor}
\usepackage{comment}
\usepackage[normalem]{ulem}

\newcommand\blfootnote[1]{%
  \begingroup
  \renewcommand\thefootnote{}\footnote{#1}%
  \addtocounter{footnote}{-1}%
  \endgroup
}

\newcommand{\oos}{\omega_s}
\newcommand{\ooi}{\omega_i}

\newcommand{\ooa}{\left( \oos,\ooi\right)}

\newcommand*\hgzero{\includegraphics[height=7pt]{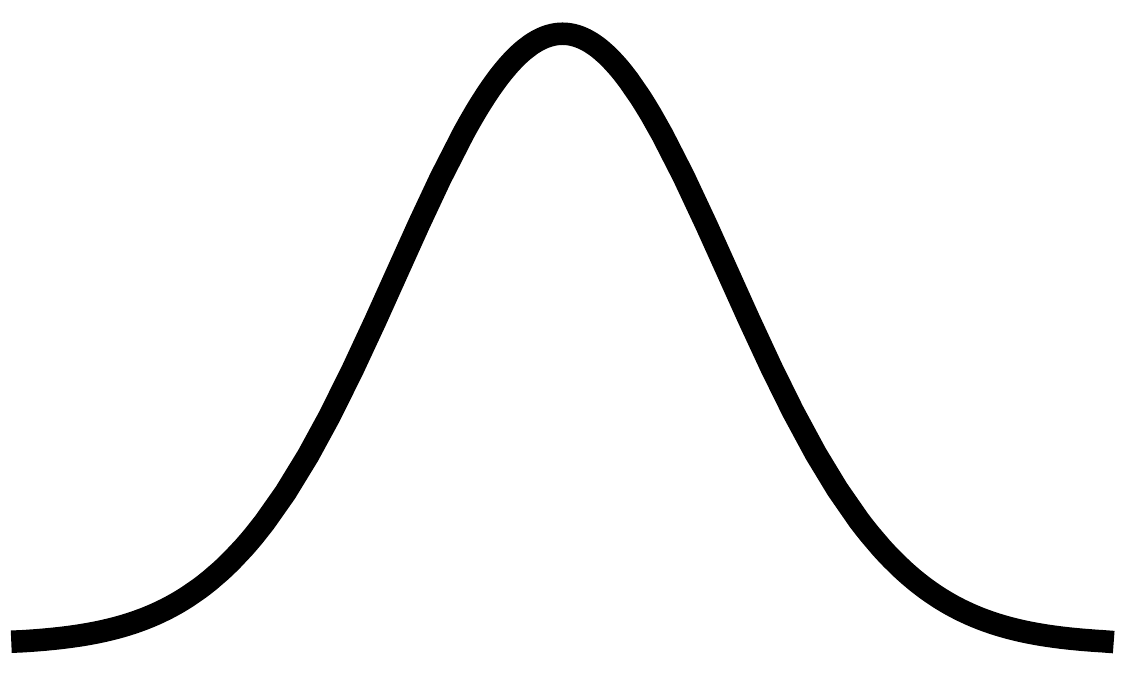}}
\newcommand*\hgone{\includegraphics[height=7pt]{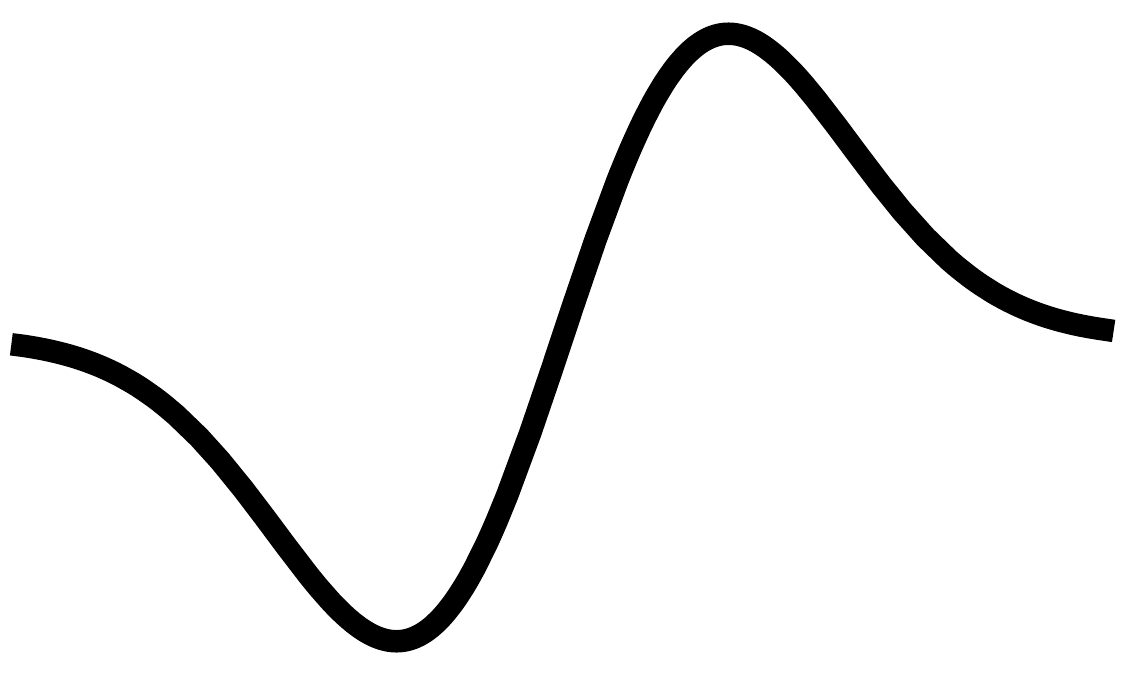}}

\begin{document}
\title{Direct generation of tailored pulse-mode entanglement}

\author{Francesco Graffitti$^{\dagger}$}
\email[Corresponding author: ]{fraccalo@gmail.com}
\blfootnote{$^{\dagger}$ These two authors contributed equally.} 
\affiliation{Institute of Photonics and Quantum Sciences, School of Engineering and Physical Sciences, Heriot-Watt University, Edinburgh EH14 4AS, UK}

\author{Peter Barrow$^{\dagger}$}
\affiliation{Institute of Photonics and Quantum Sciences, School of Engineering and Physical Sciences, Heriot-Watt University, Edinburgh EH14 4AS, UK}

\author{Alexander Pickston}
\affiliation{Institute of Photonics and Quantum Sciences, School of Engineering and Physical Sciences, Heriot-Watt University, Edinburgh EH14 4AS, UK}

\author{Agata M. Bra\'nczyk}
\affiliation{Perimeter Institute for Theoretical Physics, Waterloo, Ontario, N2L 2Y5, Canada}

\author{Alessandro Fedrizzi}
\affiliation{Institute of Photonics and Quantum Sciences, School of Engineering and Physical Sciences, Heriot-Watt University, Edinburgh EH14 4AS, UK}


\begin{abstract}
Photonic quantum technology increasingly uses frequency encoding to enable higher quantum information density and noise resilience.
Pulsed time-frequency modes (TFM) represent a unique class of spectrally encoded quantum states of light that enable a complete framework for quantum information processing.
Here, we demonstrate a technique for direct generation of entangled TFM-encoded states in single-pass, tailored downconversion processes. 
We achieve unprecedented quality in state generation---high rates, heralding efficiency and state fidelity---as characterised via highly resolved time-of-flight fibre spectroscopy and two-photon interference.
We employ this technique in a four-photon entanglement swapping scheme as a primitive for TFM-encoded quantum protocols.
\end{abstract}

\maketitle

\clearpage

Generating entanglement in intrinsically high-dimensional degrees of freedom of light, such as transverse and longitudinal spatial modes~\cite{Erhard2018,Wang285}, or time and frequency, constitutes a powerful resource for photonic quantum technologies---photons that carry more information enable more efficient protocols~\cite{PhysRevLett.88.127902,PhysRevA.88.032309}.
Time-frequency encoding is intrinsically suitable for waveguide integration and fibre transmission~\cite{PhysRevX.5.041017,Kues2017}, making it a promising choice for practical, high-dimensional quantum applications.
Quantum information can be encoded either in discrete temporal or spectral modes (namely time- and frequency-bin encoding~\cite{Kues2017,Islame1701491,Lu:18,Imany2019}) or in the spectral envelope of the single-photon wavepackets---time-frequency mode (TFM) encoding~\cite{PhysRevX.5.041017,Ansari:18integrated}. 
TFM-encoded states arise naturally in parametric downconversion (PDC) sources, as TFMs are eigenstates of the PDC process and they span an infinite-dimensional Hilbert space.
Conveniently, TFMs possess highly desirable properties:
being centred around a target wavelength makes them compatible with fibre networks, they are robust against noise~\cite{1907.02321} and chromatic dispersion~\cite{Eckstein:11}, their pulsed nature enables synchronisation and therefore multi-photon protocols and they offer intrinsically high dimensionality~\cite{Ansari:18integrated}.
Manipulation and detection of TFMs is enabled by the quantum-pulse toolbox, where sum- and difference-frequency generation are used for reshaping and projecting the quantum states~\cite{PhysRevX.5.041017,Ansari:18integrated}. However, generating entangled TFMs in a controlled way can be very challenging~\cite{PhysRevLett.94.073601,Matsudae1501223,PhysRevA.96.043822,McKinstrie:17,1907.07935}, limiting their usefulness in realistic scenarios.
Here, we overcome this problem exploiting domain-engineered nonlinear crystals~\cite{graffitti2017pure,graffitti2018independent} for generating TFM entanglement from standard ultrafast laser pulses in a single-pass PDC experiment. 
We experimentally validate this technique by benchmarking a maximally antisymmetric state at telecom wavelength with near unity fidelity, and implement a four-photon entanglement swapping scheme.
Our work complements the pulse-gate toolbox~\cite{PhysRevX.5.041017,Ansari:18integrated} for TFM quantum information processing, and establishes a standard for the generation of TFM quantum states of light while paving the way for more complex frequency encoding.

In a PDC process, a pump photon probabilistically downconverts into two photons under momentum and energy conservation. The second-order nonlinearity of a crystal mediates the process through the phase-matching function (PMF) which, together with the pump spectral profile, dictates the spectral properties of the output biphoton state in the form of its joint spectral amplitude (JSA).
The spectral entanglement between the PDC photons is quantified by the Schmidt number via Schmidt decomposition of the JSA~\cite{PhysRevA.98.053811}: a separable, unentangled JSA has a Schmidt number of $1$; higher values indicate the presence of entanglement.
Conveniently, this decomposition also provides the spectral modal structure of the PDC biphoton state. TFMs can therefore be engineered by shaping the JSA, either by modifying the pump-pulse amplitude function~\cite{Ansari:18integrated} or, as we demonstrate here, by shaping the PMF via nonlinearity engineering.
Domain-engineered crystals have been employed successfully for the generation of spectrally-pure heralded photons \cite{graffitti2017pure,graffitti2018independent}, where undesired frequency correlations are eliminated by tailoring a Gaussian nonlinearity profile. Here we extend this technique to the direct, controlled generation of custom TFM entanglement.

\begin{figure}[t!]
	\begin{center}
		\includegraphics[width=0.9\columnwidth]{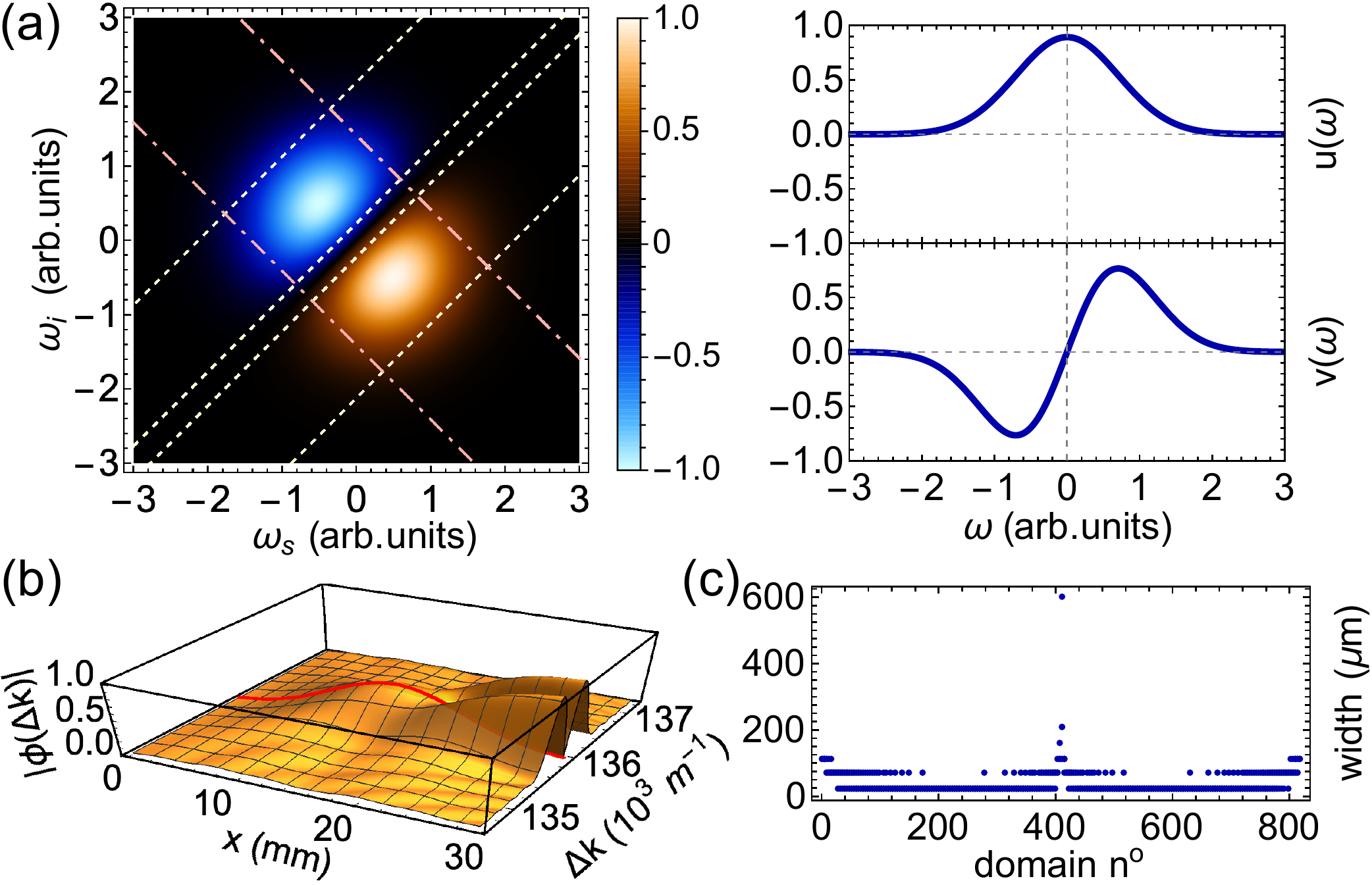}
	\end{center}
	\vspace{-1em}
	\caption{Crystal engineering.
		(a), Maximally entangled JSA (left) and corresponding TFM basis states, $u(\omega), v(\omega)$ (right). The pink dot-dashed lines and the yellow dashed lines represent the $1/\mathrm{e}$ contours of the pump function and PMF, respectively. (b), Target phase-matching function along the crystal at different momentum mismatch ($\Delta K$) values: the tracking algorithm chooses the domain orientation to track the PMF at the quasi-phase-matching condition $\Delta K =\Delta K_0$, as shown by the red trace. (c), Target crystal domain structure.}
	\label{fig:crystal}
	\vspace{-1em}
\end{figure}

We use the Hermite-Gauss modes~\cite{Ansari:18integrated} basis to encode the TFM quantum state, with the goal of generating
the maximally-entangled antisymmetric Bell-state:
\begin{equation}\begin{aligned}\label{eq:pair}
		\ket{\psi^-}_{s,i}&=\frac{1}{\sqrt{2}}\left( \ket{\hgzero}_s \ket{\hgone}_i - \ket{\hgone}_s\ket{\hgzero}_i \right)\, ,
\end{aligned}\end{equation}
where ``$s$'' (``$i$'') labels the signal (idler) photon.
The state \eqref{eq:pair} corresponds to the joint spectrum
encoded in the TFM basis states $\hgzero=u(\omega)$ and $\hgone=v(\omega)$ in Fig.~\ref{fig:crystal}(a) (see Supplemental Material for details on the biphoton spectral structure).
We use our recently-developed nonlinearity-engineering algorithm~\cite{graffitti2017pure} to shape the PMF ($\phi(x,\Delta k)$) as a first-order Hermite-Gauss function. 
We design a $30$~mm potassium titanyl phosphate (KTP) crystal for symmetric group-velocity matching with a $1.3$~ps laser pulse~\cite{PhysRevA.98.053811}. The fundamental domain width is $\sim23.1$ $\upmu$m, equal to the coherence length of a $775$~nm pump downconverted into two $1550$~nm photons. Our algorithm chooses the ferroelectric orientation of individual domains to track a target PMF along the field propagation in the crystal (see Supplemental Material for details on the algorithm). Fig.~\ref{fig:crystal}(b,c) show the resulting PMF ($\phi(\Delta k)$ at $x=30$~mm) and the required crystal domain configuration. 

\begin{figure}[h]
	\begin{center}
		\includegraphics[width=0.9\columnwidth]{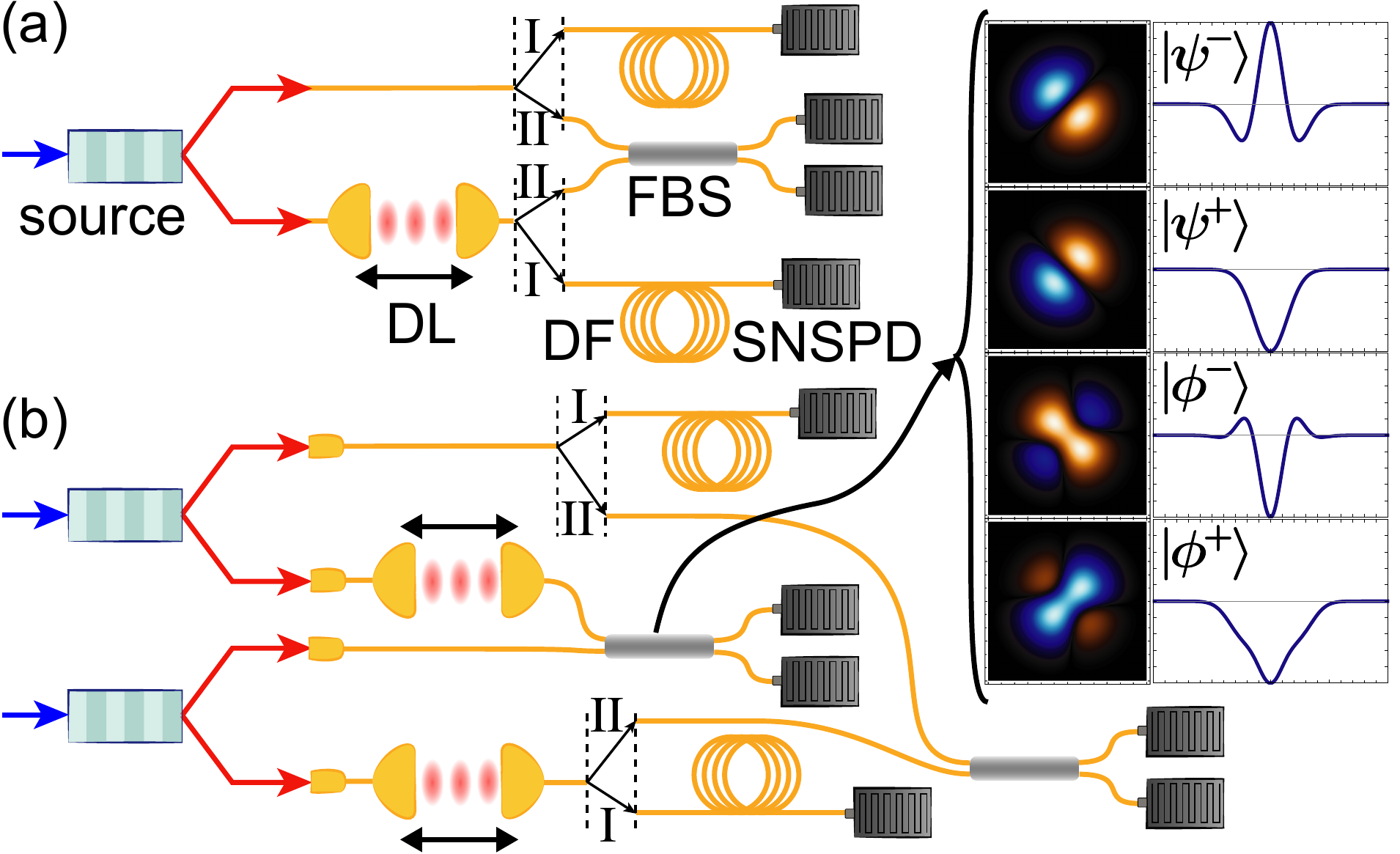}
	\end{center}
	\vspace{-1em}
	\caption{Experimental setup.
		(a), Biphoton state characterisation:
		Joint-spectrum reconstruction via dispersive fibre (DF) time-of-flight spectroscopy (modes I) and two-photon interference in a fibre beam-splitter (FBS) (modes II).
		(b), Entanglement swapping setup: Successful entanglement swapping is heralded by a coincidence detection of the photons after the FBS. The swapped state is again verified via fibre spectroscopy (modes I) and two-photon interference (modes II). We note that a setup similar to (modes I) has been used to investigate the spectral properties of two-photon interference~\cite{Jin:15}. The panel on the right shows the four possible Bell-state projections at the BS, and the corresponding interference pattern.
	}
	\label{fig:setup}
	\vspace{-1em}
\end{figure}

\begin{figure*}[t]
	\begin{center}
		\includegraphics[width=1.9\columnwidth]{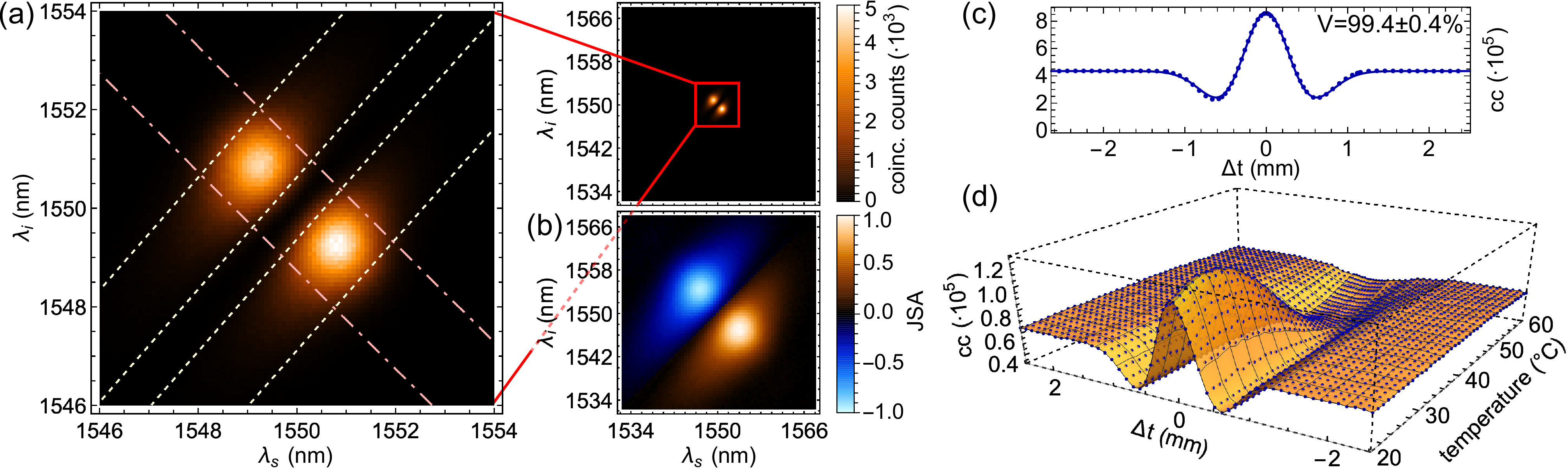}
	\end{center}  
	\vspace{-1em}
	\caption{Single source characterisation.
		(a), 
		Measured JSI (right) and zoom on a reduced, $8$~nm spectral range (left) to show its main features. 
		The dot-dashed pink lines and yellow dashed lines represent the $1/\mathrm{e}$ contours of the sech$^2$ pump function, and the PMF calculated from the crystal domain structure, respectively.
		(b), Reconstructed effective JSA.
		(c), Signal-idler interference pattern.
		(d), Signal-idler interference varying crystal temperature. The interference visibility has a maximum at $25$ degrees, while no antibunching occurs above $39$ degrees.
		Error bars assuming Poissonian counting statistics are smaller than the symbol size.}
	\label{fig:twofold}
	\vspace{-1em}
\end{figure*}

The designed crystal was manufactured commercially by \emph{Raicol Ltd}. 
We set up a collinear PDC source~\cite{graffitti2018independent}, 
where a $80$~MHz, pulsed Ti:Sapphire laser is focused into the tailored KTP crystal to create orthogonally-polarised photon pairs via type-II PDC. 
The photons are loosely filtered with a bandpass filter ($\sim3$ times broader than the PDC photons' bandwidth). A polarising beamsplitter separates the PDC photons before they are coupled into single-mode fibres.
We measured a source brightness of $\sim 4$~KHz/mW photon pairs with a symmetric heralding efficiency $>60\%$, a reasonable trade-off achieved optimising the pump, signal and idler focusing conditions~\cite{graffitti2018independent}.

A full characterisation of the biphoton quantum state could be obtained via quantum state tomography in the TFM basis, which requires projective measurements onto three mutually unbiased bases using cascades of tailored nonlinear processes~\cite{Huang:13,PhysRevLett.120.213601,Reddy:18}, or by reconstructing the JSA including its phase, which assumes a pure biphoton state and involves complex interferometric techniques~\cite{Jizan:16,1809.03727,1901.08849}. We instead characterise the PDC state using an indirect approach that exploits joint spectral intensity (JSI) reconstruction via dispersive fibre spectroscopy~\cite{Avenhaus:09} and two-photon interference~\cite{graffitti2018independent} to infer information on the populations and the entanglement of the quantum state, respectively.

The setup for the JSI reconstruction is shown in Fig.~\ref{fig:setup}(a) (modes I). Each photon is sent through a $\sim20$~km single-mode fibre to convert spectral to temporal information exploiting the fibre dispersion of $\sim18$~ps/km/nm at $1550$~nm. 
The photons are then detected with superconductive nanowire single photon detectors (SNSPD), with $\sim 80$\% detection efficiency and $<50$~ps timing jitter. Arrival times are recorded as time tags by a \textit{Picoquant HydraHarp} in $1$~ps bins for offline processing.
We collected $\sim2.8 \times 10^6$ two-photon coincidence counts with respect to a third reference SNSPD in $24$~hours.
We reconstruct the JSI over a $36$~nm spectral range, $\sim12$~times larger than the PDC photons' bandwidth, to ensure reliable estimation of the JSI properties~\cite{PhysRevA.98.053811}. The results are shown in Fig.~\ref{fig:twofold}(a).
The overlay contours show the theoretical pump spectrum and the expected PMF (assuming the ideal crystal domain structure and a sech$^2$ pump function). There is excellent correspondence between the theoretical target and the measured JSI, which faithfully reproduces not only the two main peaks but also the spectral bandwidth.

The two-photon interference setup is shown in Fig.~\ref{fig:setup}(a), modes II: the interference pattern is measured by delaying one photon with respect to the other before they interfere in a fibre BS. 
While two-photon interference is typically destructive and, for PDC photons, exhibits a characteristic triangular or Gaussian ``dip''~\cite{graffitti2018independent}, antibunching at the BS can occur whenever the biphoton state is at least partially antisymmetric under particle exchange: more antisymmetry results in more antibunching~\cite{Fedrizzi_2009} (see Supplemental Material for proof).
Remarkably, for a biphoton state that is separable in all other DOFs, antibunching corresponds to entanglement in the biphoton spectrum \cite{douce2013direct}.
We use this result to verify TMF entanglement in our generated state.
We show the experimental data in Fig.~\ref{fig:twofold}(c): the fitted interference visibility is equal to $99.4\pm0.4\%$, certifying a high degree of spectral entanglement and purity of the PDC biphoton state (the fitting function is given in Supplemental Material).

While the JSI reconstruction doesn't contain any phase information, we can exploit our knowledge of the purity and the now verified antisymmetry of the biphoton wavefunction to infer an effective JSA of the PDC state.
Specifically, to guarantee the overall antisymmetry of the quantum state, we impose a $e^{i \pi}$ sign shift between the two peaks of the square root of the measured JSI. We obtain the JSA depicted in Fig.~\ref{fig:twofold}(b), that qualitatively matches the theoretical JSA shown in Fig.~\ref{fig:crystal}(a).
We can now quantify the entanglement and spectral structure of our biphoton state via Schmidt decomposition. 
Since the biphoton $\ket{\psi^-}$ state in \eqref{eq:pair} is composed of two, equally weighted TFM basis states, the corresponding Schmidt number is equal to 2.
We extract an experimental Schmidt number equal to $2.026\pm0.001$ from the reconstructed JSA, in excellent agreement with the theoretically expected value (details on the JSI reconstruction and error estimation are discussed in Supplemental Material). 

Small variations in the crystal domain widths can be introduced by changing the crystals temperature. This results in a shift of the PMF in the $\ooa$ plane, producing frequency non-degenerate photons and therefore compromising the antisymmetry of the biphoton wavefunction.
Surprisingly, this doesn't affect the Schmidt number of the quantum state: the biphoton state \eqref{eq:pair} remains intact, but the signal and idler TFMs will be centred around different frequencies. This enables the capability of switching between an antisymmetric state to a non-antisymmetric one without spoiling the spectral modal structure.
We observe the biphoton antisymmetry-breaking by performing interference scans at different temperatures, from $20$ to $60$ degrees at $1$ degree intervals. 
We show the results in Fig. \ref{fig:twofold}(d):
anti-bunching (and therefore antisymmetry) is maximal for perfectly degenerate PDC and it reduces as we tune away from degeneracy, until no anti-bunching occurs above a certain centre-frequency offset, as expected from theory.

Multiphoton protocols using TFMs will require the ability to interfere and swap independently generated TFM-encoded photons.
While a generalised entanglement swapping for TFM has been proposed, it relies on a nonlinear process between two single photons and therefore has very low success probability~\cite{PhysRevA.98.023836}.
Here we instead implement the standard entanglement swapping scheme with the setup shown in Fig. \ref{fig:setup}(b).
Two entangled $\ket{\psi^-}$ states are produced via two independent engineered TFM-entangled pair sources. The overall four-photon state can be written as
$1/2(\ket{\phi^+}\ket{\phi^+}+\ket{\phi^-}\ket{\phi^-}+\ket{\psi^+}\ket{\psi^+}-\ket{\psi^-}\ket{\psi^-})$
, a coherent sum of the four Bell states:
\begin{equation}
	\begin{aligned}
		\begin{matrix}\ket{\psi^\pm}\\\ket{\phi^\pm}\end{matrix}=\frac{1}{\sqrt{2}}
		\begin{matrix} \ket{\hgzero} \ket{\hgone} \pm \ket{\hgone}\ket{\hgzero}\\ 
			\ket{\hgzero} \ket{\hgzero} \pm \ket{\hgone}\ket{\hgone}\end{matrix}
	\end{aligned}
	\label{eq:bellbasis}
\end{equation}
The joint spectra for all four Bell states and the corresponding interference patterns are shown in the inset of Fig.~\ref{fig:setup}(b): perfect antibunching at the BS occurs only for the singlet state, while the triplet states bunch due to the symmetry of their wavefunctions.
We use this to discern a successful projection on $\ket{\psi^-}$ from all the other outcomes: a two-photon coincidence detection at the two BS outputs corresponds to a projection on the singlet state and heralds swapping of the TFM $\ket{\psi^-}$ state from the two original photon pairs to the two remaining non-interacting photons.

\begin{figure}[t!]
	\begin{center}
		\includegraphics[width=0.9\columnwidth]{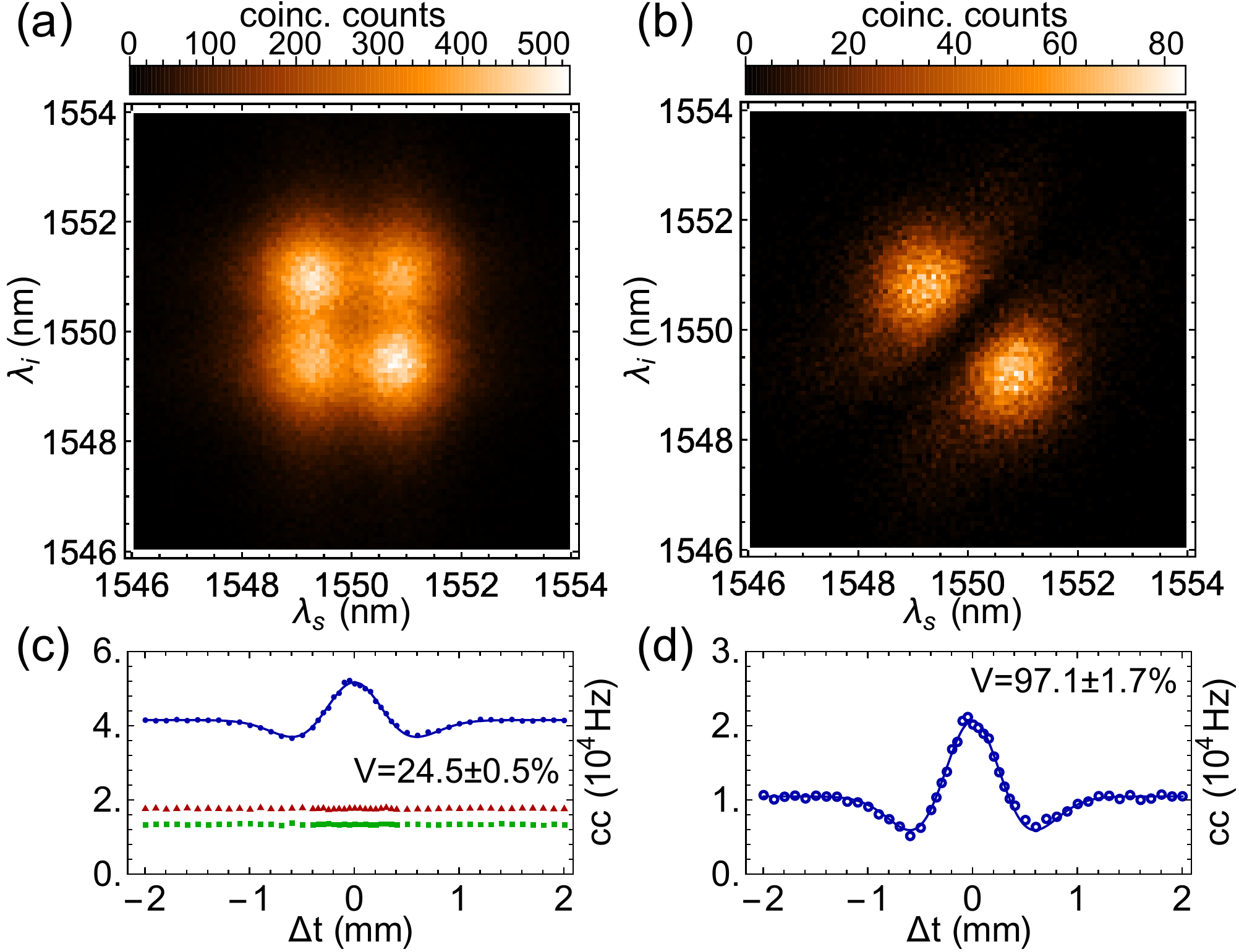}
	\end{center}
	\vspace{-1em}
	\caption{Entanglement swapping results.
		(a), JSI reconstruction of the fully-mixed state. The two peaks on the antidiagonal have higher count rates because the contribution from the $\ket{\psi^-}$ state is counted twice in the 3-fold coincidences.
		(b), JSI reconstruction of the entanglement-swapped state. We display a $8$~nm spectral range to highlight the main JSI features.
		(c), Two-photon interference without signal error correction. The blue data points are the 4-fold coincidence counts detected by the SNSPDs when both the sources are active, while red triangles and green squares are the error signals measured by alternately blocking one of the two sources.
		(d), Two-photon interference with higher-order emissions corrected.
	}
	\label{fig:fourfold}
	\vspace{-1em}
\end{figure}

We benchmark the state obtained after entanglement swapping via fibre spectroscopy and two-photon interference, as shown in Fig.~\ref{fig:setup}(b), modes I and II, respectively. The JSI of the swapped $\ket{\psi^-}$ state is again measured by sending the two photons through a pair of $20$~km fibres.
We collect $670$k 3-fold and $\sim46$k 4-fold coincident counts in 72 hours of integration time.
In Fig.~\ref{fig:fourfold}(a) we show the measured joint spectrum of the two-photon state without post-selection, heralded by either one or two detection events after the BS. 
We observe 4 peaks, arising from a mixture of the four equally weighted Bell state JSAs (see Fig.~\ref{fig:setup}(b)).
When we instead record 4-fold coincident counts, we measure the spectrum of the swapped $\ket{\psi^-}$ biphoton state, recovering the two main peaks on the JSI's diagonal (Fig.~\ref{fig:fourfold}(b)).

We then measure the interference of the swapped state. Because the probability of generating photon pairs independently equals that of a double-pair emission in each source, the maximal theoretical interference visibility is $25\%$---not a fundamental limitation, it only occurs when both photons of two PDC pairs are interfered, which is not required for e.g. repeater protocols. We obtain an interference visibility of $24.5\pm0.5\%$, as shown in Fig.~\ref{fig:fourfold}(c). We subtract the multi-photon background by recording the counts when either of the two photon sources are blocked. The corrected interference pattern in Fig.~\ref{fig:fourfold}(d) yields an interference visibility of $97.1\pm1.7\%$, certifying success of the TFM entanglement swapping protocol.

We can now reconstruct the JSA of the swapped state as described earlier, calculating a Schmidt number of $2.15\pm0.01$---slightly higher than for the single-source scenario, as expected due to inevitable discrepancies between independent sources that affect the interference quality.

We have demonstrated the first instance of TFM entanglement generation enabled by nonlinearity engineering, achieving high generation rates, heralding efficiency and spectral entanglement.
Due to its simplicity and quality, we expect this technique to be used in a host of different quantum information tasks.
The flexibility in tailoring the PMF lends itself to the generation of high-dimensional TFM entanglement: not only can one use higher-order Hermite-Gaussian PMFs to up-scale to qudits~\cite{Ansari:18integrated}, but one can also aim at different PMF shapes for targeting specific applications, such as frequency multiplexing~\cite{1907.10355}. 
The same nonlinearity engineering technique can be used in asymmetric group-velocity matching condition~\cite{PhysRevA.98.053811} to generate pure, TFM-encoded single photons, as well as to implement mode filtering and TFM-projective measurements in a quantum pulse gate scheme, complementing the TFM framework based on pump spectral-shaping \cite{PhysRevX.5.041017,Ansari:18integrated}.
Finally, the ability to customise biphoton spectra could be useful for multi-photon quantum metrology applications in which measurement precision depends on the shape and steepness of the interference pattern~\cite{Lyonseaap9416}.

\smallskip
We thank J.~Leach and M.~Malik for loan of equipment, and D.~Kundys and M.~Proietti for useful discussions.

\smallskip
This work was supported by the UK Engineering and Physical Sciences Research Council (grant number EP/N002962/1). FG acknowledges studentship funding from EPSRC under grant no. EP/L015110/1. Research at Perimeter Institute is supported by the Government of Canada through Industry Canada and by the Province of Ontario through the Ministry of Research and Innovation. Natural Sciences and Engineering Research Council of Canada (RGPIN-2016-04135). 

\smallskip
\noindent See Supplemental Material for supporting content.

%

\end{document}


\title{Direct generation of tailored pulse-mode entanglement - Supplemental Material}

\author{Francesco Graffitti}
\affiliation{Institute of Photonics and Quantum Sciences, School of Engineering and Physical Sciences, Heriot-Watt University, Edinburgh EH14 4AS, UK}

\author{Peter Barrow}
\affiliation{Institute of Photonics and Quantum Sciences, School of Engineering and Physical Sciences, Heriot-Watt University, Edinburgh EH14 4AS, UK}

\author{Alexander Pickston}
\affiliation{Institute of Photonics and Quantum Sciences, School of Engineering and Physical Sciences, Heriot-Watt University, Edinburgh EH14 4AS, UK}

\author{Agata M. Bra\'nczyk}
\affiliation{Perimeter Institute for Theoretical Physics, Waterloo, Ontario, N2L 2Y5, Canada}

\author{Alessandro Fedrizzi}
\affiliation{Institute of Photonics and Quantum Sciences, School of Engineering and Physical Sciences, Heriot-Watt University, Edinburgh EH14 4AS, UK}


\begin{abstract}
This document provides Supplemental Material to ``Direct generation of tailored pulse-mode entanglement''.
The document is structured as follows: In Section 1 we introduce the theoretical JSA and its components; In Section 2 we discuss the details of the nonlinearity engineering technique; In Section 3 we calculate the exact correspondence between the symmetry of the JSA and the interference visibility; In Section 4 we give some details on the JSI reconstruction.
\end{abstract}

\maketitle

\renewcommand{\theequation}{S\arabic{equation}}
\renewcommand{\thefigure}{S\arabic{figure}}
\renewcommand{\thetable}{S\arabic{table}}
\renewcommand{\thesubsection}{S\Roman{subsection}}
\setcounter{equation}{0}
\setcounter{figure}{0}
\setcounter{table}{0}
\setcounter{section}{0}

\section{1. Theoretical JSA and spectral modes} 
The JSA in Fig. 1(a) of the main text is obtained by combining a Gaussian pump with an antisymmetric PMF shaped as the first order Hermite-Gaussian function multiplied by a Gaussian envelope:
\begin{equation}\begin{aligned}
		\alpha \ooa &= e^{-\frac{\left(\omega_s+\omega_i\right)^2}{2\sigma^2}}\\
		\phi\ooa &= e^{-\frac{\left(\omega_s-\omega_i\right)^2}{2\sigma^2}} \left(\omega_s-\omega_i\right)\, .
\end{aligned}\end{equation}
The corresponding PDC first-order emission biphoton state reads:
\begin{equation}\begin{aligned}
		\ket{\psi^-\ooa}_{s,i}= &\iint d\oos d\ooi \alpha\ooa \phi\ooa a_s^\dagger(\oos)a_i^\dagger(\ooi)\ket{0}_{s,i}\\
		=&\iint d\oos d\ooi \exp{\left[-\frac{\omega_s^2+\omega_i^2}{\sigma^2}\right]}  \left(\omega_s-\omega_i\right)\\ & a_s^\dagger(\oos)a_i^\dagger(\ooi)\ket{0}_{s,i} \, ,
		\label{eq:pdcstate}\end{aligned}\end{equation}
where we are omitting the wavefunction normalisation.
\eqref{eq:pdcstate} can be decomposed into the convex sum of a set of orthonormal one-variable function by performing the Schmidt decomposition:
\begin{equation}\begin{aligned}
		&\ket{u(\omega_j)}_j \equiv \ket{\hgzero}_j = \int d\omega_j\exp{\left[-\frac{\omega_j^2}{\sigma^2}\right]}a_j^\dagger(\ooi)\ket{0}_{j}\\
		&\ket{v(\omega_j)}_j \equiv \ket{\hgone}_j = \int d\omega_j  \exp{\left[-\frac{\omega_j^2}{\sigma^2}\right]}\omega_j a_j^\dagger(\ooi)\ket{0}_{j}\, ,
		\label{eq:schmidt}\end{aligned}\end{equation}
with $j=s,i$. The shape of the two spectral modes is shown in Fig. 1(a). Following from \eqref{eq:schmidt}, the biphoton state can be therefore written as follows:
\begin{equation}\begin{aligned}
		\ket{\psi^-\ooa}_{s,i}&= \frac{1}{\sqrt{2}}\left( \ket{u(\oos)}_s \ket{v(\ooi)}_i - \ket{v(\oos)}_s\ket{u(\ooi)}_i \right)\\
		&=\frac{1}{\sqrt{2}}\left( \ket{\hgzero}_s \ket{\hgone}_i - \ket{\hgone}_s\ket{\hgzero}_i \right)\, .
		\label{eq:pdcsinglet}\end{aligned}\end{equation}
The state in \eqref{eq:pdcsinglet} is a maximally entangled singlet state in the spectral-temporal mode basis.

The expected two-photon interference pattern corresponding to this state can be calculated as described in~\cite{1711.00080}, and has the form:
\begin{equation}
	\begin{aligned}
		p_{\text{cc}}(\Delta t)=\frac{1}{2}-\frac{1}{4} e^{-\frac{1}{4}\sigma^2\Delta t^2}(\sigma^2\Delta t^2-2) \, ,
	\end{aligned}
	\label{eq:HOM}
\end{equation}
where $p_{cc}$ is the coincidence-count probability after  interference, $\sigma$ depends on the biphoton bandwidth and $\Delta t$ is the relative arrival time of the two photons at the BS. We use \eqref{eq:HOM} (with an additional visibility scaling factor) to fit the two-photon interference data.

\section{2. Engineering algorithm}
We use the domain engineering technique introduced in \cite{graffitti2017pure} to shape the PMF of the crystal to an almost-arbitrary function.
For generating the two-photon entangled state described in \eqref{eq:pdcsinglet}, we need a PMF ($\Phi$) of the form:
\begin{equation}
	\Phi(\Delta k)=\exp \left[- \frac{\sigma^2}{2}(\Delta k-\Delta k_0)^2\right]  (\Delta k-\Delta k_0)\, ,
	\label{eq:pmf}    
\end{equation}
where $\sigma$ is the parameter determining the PMF's width, $\Delta k$ is the momentum mismatch, $\Delta k_0=\pi/\ell$ is the quasi-phase-matching momentum, being $\ell$ the coherence length of the process ($2\ell$ corresponds to the poling period of a standard periodically poled crystal).
An inverse Fourier transform of \eqref{eq:pmf} gives the target nonlinearity profile along the crystal:
\begin{equation}\begin{aligned}
		g_{\text{target}}&(x)=\ \mathcal{FT}^{-1} \left[\Phi(\Delta k) \right]=\\
		&i\ \exp \left[-\frac{(x-\frac{L}{2})^2}{2 \sigma^2}\right] \exp \left[i \Delta k_0 x\right] \left(x - \frac{L}{2}\right) \, ,
		\label{eq:gtarget}\end{aligned}\end{equation}
with $L$ the crystal length, and where we have omitted the multiplicative constant. 
By integrating $g_{\text{target}}(x)$ along the longitudinal direction of the crystal, we obtain the target phase-matching function that our algorithm needs to track:
\begin{equation}\begin{aligned}
		\Phi_{\text{target}} \left(x,\Delta k=\Delta k_0\right) = - i \left. \int_0^x  g_{\text{target}}(x') e^{i \Delta k x'} dx'\right\rvert_{\Delta k_0}=\\
		- i\ \frac{2 \sqrt{e}}{\pi \sigma}\exp \left[ - \frac{L^2+4 x^2}{8\sigma^2}\right] \left( \exp \left[\frac{x^2}{2\sigma^2}\right] - \exp \left[\frac{L \ x}{2 \sigma^2}\right] \right) \sigma^2 \, ,
		\label{eq:pmftarget}\end{aligned}\end{equation}
where the prefactor is chosen for matching the maximum function's slope allowed by the field tracking algorithm in order to maximise the photon pairs production \cite{graffitti2017pure}. The parameters chosen for this experiment are $L=30$mm and $\sigma = L/5$. Fig. 1(b) in the main text shows the tracking function described in \eqref{eq:pmftarget} and the corresponding overall phase-matching function along the crystal.

\section{3. Correspondence between JSA antisymmetry and interference antibunching}
Any JSA can be decomposed in its symmetric and antisymmetric parts as follows:
\begin{equation}\begin{aligned}
		&f\ooa = \gamma \frac{f\ooa + f(\ooi,\oos)}{2} + \delta \frac{f\ooa - f(\ooi,\oos)}{2} \\
		&f\ooa = \gamma f_s\ooa+ \delta f_a\ooa\, ,
		\label{eq:deco}\end{aligned}\end{equation}
where $f_s\ooa=f_s(\ooi,\oos)$ and $f_a\ooa=-f_a(\ooi,\oos)$. 
Imposing a normalisation condition for $f_s$ and $f_a$:
\begin{equation}\begin{aligned}
		& \iint d\oos d\ooi \abs{f_s\ooa}^2 = \iint d\oos d\ooi \abs{f_a\ooa}^2 =1\, ,
		\label{eq:norm}\end{aligned}\end{equation}
implies that $\gamma$ and $\delta$ need to satisfy the following condition $\abs{\gamma}^2 + \abs{\delta}^2 = 1$.
The probability of having coincident counts after the BS (i.e. antibunching) reads: 
\begin{equation}\begin{aligned}
		p_{\text{cc}}(\Delta t)= \frac{1}{2} - \frac{1}{2} \iint d\oos d\ooi f^*\ooa f(\ooi,\oos) e^{i (\ooi - \oos) \Delta t} \, .
		\label{eq:hom_general}\end{aligned}\end{equation}
We now replace the JSA with its decomposition in symmetric and antisymmetric parts, and consider the photons arriving at the same time at the BS ($\Delta t = 0$):
\begin{equation}\begin{aligned}
		p_{\text{cc}}(0)
		=&\frac{1}{2} - \frac{1}{2} 
		\iint d\oos d\ooi \\
		&(\abs{\gamma}^2 \abs{f_s\ooa}^2 -\gamma^* \delta f_s^*\ooa f_a\ooa\\ &+\gamma \delta^* f_a^*\ooa f_s\ooa -\abs{\delta}^2\left.\abs{f_a\ooa}^2\right)\, .
\end{aligned}\end{equation}
The integral of the mixed terms is equal to zero because the overall product of  $f_s$ and $f_a$ is antisymmetric, and considering the normalisation conditions in \eqref{eq:norm} the coincidence probability reads:
\begin{equation}\begin{aligned}
		p_{\text{cc}}(0)=\frac{1}{2} - \frac{1}{2} (\abs{\gamma}^2 - \abs{\delta}^2) = 1 - \abs{\gamma}^2 = \abs{\delta}^2
		\, .
\end{aligned}\end{equation}

\section{4. JSI reconstruction and error estimation}
The measured JSIs are $12250 \times 12250$ matrices, where each bin has a size of $1\times1$~ps, corresponding to the timing logic's resolution.
We calibrate our dispersive-fibre spectrometer with a reference signal with respect to a commercial single-photon CCD spectrometer, obtaining a scaling factor of $\sim2.94$~pm/ps (centred around $1550$~nm). This corresponds to a total measurement spectral range of $\sim36$~nm.
We down-sample the JSIs to $40\times40$~ps bins for reducing the sparsity of the data and computing the singular value decomposition (as numerical implementation of the Schmidt decomposition). 
The error on the extracted Schmidt numbers represents $3\sigma$ statistical confidence regions obtained via Monte-Carlo re-sampling ($10$k runs of the algorithm) assuming a Poissonian statistics on the coincident counts distribution.

%